# Optical trap potential control in N-type four level atoms by femtosecond Gaussian pulses

Subhadeep Chakraborty and Amarendra K. Sarma*
Department of Physics, Indian Institute of Technology Guwahati, Guwahati-781039, Assam, India.
*Electronic address: aksarma@iitg.ernet.in

In this work we present a scheme to control the optical dipole trap potential in an N-type four-level atomic system by using chirped femtosecond Gaussian pulses. The spatial size of the trap can be well controlled by tuning the beam waist of the Gaussian pulse and the detuning frequency. The trapping potential splits with increasing Rabi frequency about the center of the trap, a behavior analogous to the one observed experimentally in the context of trapping of nanoparticles with femtosecond pulses. An attempt is made to explain the physics behind this phenomenon by studying the spatial probability distribution of the atomic populations.

Manipulation and trapping of neutral atoms using laser pulses is continued to be a topic of considerable interest in atomic and optical physics [1]. In recent years, owing to the tremendous technological progress in the generation of femtosecond and attosecond laser pulses study of optical force on atoms and molecules is getting a tremendous boost [2-7]. Optical force is now exploited in as diverse areas as atom optics [8], Bose-Einstein condensation (BEC) [9,10] and quantum information [11]. In the context of BEC, it is now well established that many technical problems in realizing BEC could be avoided if optical trap based on optical dipole force is used [10]. It is worthwhile to note that optical trapping has been studied quite extensively in the context of trapping and manipulating nano-particles also [12]. In a recent experimental study, based on the interaction of femtosecond laser pulses and induced polarization, it is shown how optical trap can be constructed and manipulated to trap gold nanoparticles [12]. It is shown that the trap potential is split into two when the incident peak laser power exceeds a threshold level. In atomic physics, the peak-laser power is related to the so-called Rabi frequency. In this brief report we report a remarkable analogous splitting of optical potential in a N-type four level atomic system.

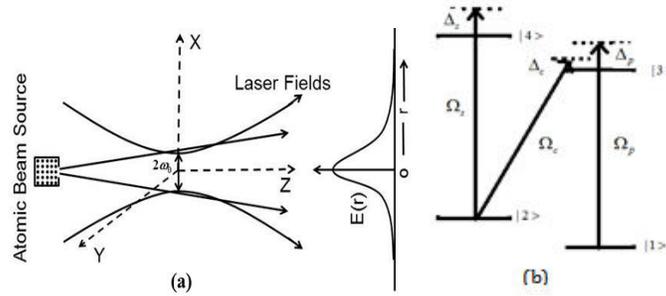

FIG. 1. (a) Schematic diagram of an atomic beam co-propagating with laser fields (b) N-type four level atomic system

Our analysis is based on the scheme depicted in Fig. 1. The four level N-type atomic system is depicted in Fig.1 (b). Each of the three laser pulses interacts only with a pair of energy levels. The signal, coupling, and pump fields drive the atomic transitions $|2> - |4>$, $|2> - |3>$, and $|1> - |3>$, respectively. The transitions $|1> - |2>$, and $|3> - |4>$ are dipole forbidden. $\Omega_s$, $\Omega_c$, and $\Omega_p$, are the respective Rabi frequencies for the signal, coupling and pump fields. The general expression for time dependent Rabi frequencies is given by:

$$\Omega_{p,c,s} = A_{p,c,s}(t)\cos(\omega_{p,c,s}t + \chi_{p,c,s}t^2 + \omega_{D_{p,c,s}}) \tag{1}$$

Here, $\omega_{p,c,s}$ are the carrier frequencies and $\chi_{p,c,s}$ are the chirping rates of the pump, the coupling and the signal pulses respectively. $\omega_{D_{p,c,s}} = \vec{k}_{p,c,s} \cdot \vec{v}$ represents the frequency detuning of the atom moving with velocity $\vec{v}$ due to Doppler effect. $A_{p,c,s}(t)$ refers to the pulse envelope and are assumed to be Gaussian function in form:

$$A_{p,c,s}(t) = \hat{\eta}\, \Omega_{p0,c0,s0} exp\left(-\frac{r^2}{\omega_0^2} - \frac{t^2}{\tau^2}\right) \qquad (2)$$

Here $\hat{\eta}$ and $\omega_0$ are respectively the polarization vector and the beam waist of the pulse. $\tau$ is the temporal width of the pulse, related to the FWHM of the laser pulse by: $\tau_p = 1.177\tau$. The time independent Rabi frequencies are defined as: $\Omega_{p0,c0,s0} = \mu_{p,c,s} E_0/\hbar$, where $E_0$ is the peak amplitude of the electric fields and $\mu_{p,c,s}$ are the dipole moments corresponding to different transitions. The detuning frequency to be used in the rest of the work is defined as: $\Delta_p = \omega_p - \omega_{31}$, $\Delta_c = \omega_c - \omega_{32}$ and $\Delta_s = \omega_s - \omega_{42}$. For simplicity we consider that the Rabi frequency, the chirping rate and the detuning frequency are equal for all the pulses i.e. $\Omega_{p0} = \Omega_{c0} = \Omega_{s0} = \Omega_0$, $\chi_p = \chi_c = \chi_s = \chi$ and $\Delta_p = \Delta_c = \Delta_s = \Delta$. It may be noted that this model is studied in the context of electromagnetic induced absorption [13], electromagnetic induced transparency (EIT) [14], quantum interference [15] and quantum coherence and population trapping [16] etc. We have analyzed the optical force and optical potential by numerically solving the appropriate density matrix equations. Based on an approach of Ehrenfest's theorem [4] the transverse component of optical dipole force is given by:

$$F_t = -(2r\hbar/\omega_0^2)\left[(\rho_{42} + \rho_{24})\Omega_s(t) + (\rho_{32} + \rho_{23})\Omega_c(t) + (\rho_{31} + \rho_{13})\Omega_p(t)\right] \qquad (3)$$

Due to the conservative character, the force can be derived from a potential. This optical potential can be defined by the following equation:

$$U = -\hbar\left[(\rho_{42} + \rho_{24})\Omega_s(t) + (\rho_{32} + \rho_{23})\Omega_c(t) + (\rho_{31} + \rho_{13})\Omega_p(t)\right] \qquad (4)$$

Here $\rho_{ij}$ refers to the density matrix elements [15]. Based on $^{87}$Rb hyperfine structure we set the states as: $|1\rangle = |5^2S_{1/2}, F=1\rangle$, $|2\rangle = |5^2S_{1/2}, F=2\rangle$, $|3\rangle = |5^2P_{3/2}, F'=1\rangle$ and $|4\rangle = |5^2P_{3/2}, F'=3\rangle$ [15]. The following typical parameters are used for simulation: $\omega_{42} = 2.414176$ rad/fs, $\omega_{31} = 2.369459$ rad/fs, $\omega_{32} = 2.369416$ rad/fs and $M = 1.443 \times 10^{-25} Kg$ [17]. It should be noted that the scheme presented here is independent of the chosen numerical parameters and may be applicable to some other real atoms also. The other numerical parameters are $v = 100\ m/s$ and $\tau = 15.5\ fs$.

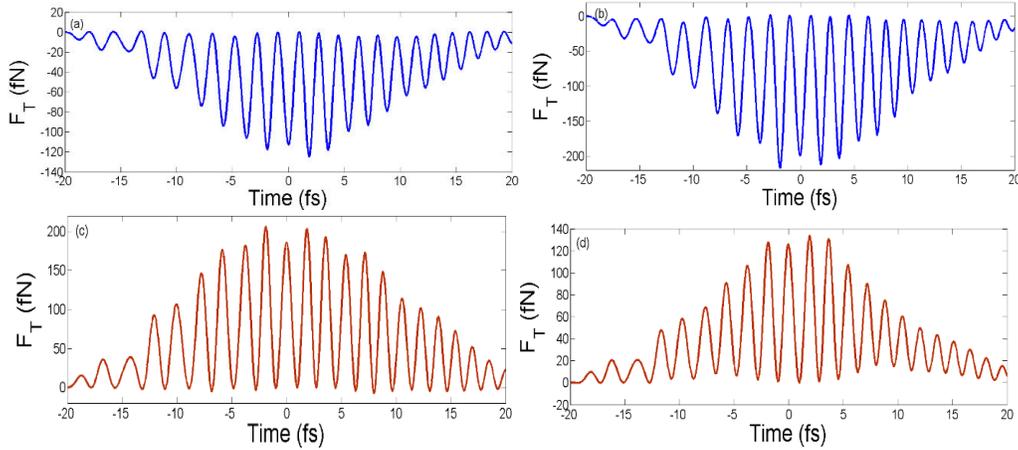

FIG. 2. Temporal evolution of transverse optical dipole force with $r = 0.3\ \mu m$, $\omega_0 = 0.5\mu m$, $\Delta = -0.7 rad/fs$, $\Omega_0 = 1.0\ rad/fs$, $\chi = 0.01 fs^{-2}$ (a) $\rho_{11}(-\infty) = 1$ (b) $\rho_{22}(-\infty) = 1$ (c) $\rho_{33}(-\infty) = 1$ (d) $\rho_{44}(-\infty) = 1$.

Fig. 2 depicts the temporal evolution of transverse optical force for different initial states, under the same pulse parameters. In Fig. 2(a) and 2(b) the atom is initially prepared in the two ground levels |1> and |2> respectively. The transverse optical force is negative and directed towards the z-axis. The atoms experience a net attractive force which leads to the focusing of the atoms around z=0 axis. On the other hand Fig. 2(c) and 2 (d) refers to the case when the atom is initially prepared in the two excited levels, respectively |3> and |4>, under the same pulse parameters as in 2(a)-(b). Here the optical force is positive and directed away from the z-axis resulting in defocusing of the atom. The optical force changes from focusing to defocusing in the transverse direction as the initial population changes from the ground to the excited levels of the atom. So focusing and defocusing of atom using transverse optical force can be well controlled by properly choosing the initial state of the atom, keeping the pulse parameters unchanged. For the chosen numerical parameters of the scheme, acceleration of the atom, about 10 orders of magnitude higher than Earth's gravitational acceleration g is achieved. It should also be noted that the magnitude of the transverse optical force is greater when the atom is initially prepared in intermediate levels |2> and |3> than |1> and |4>. So one can easily distinguish the initial state of an N type four level atom just by measuring the magnitude and direction of transverse optical force.

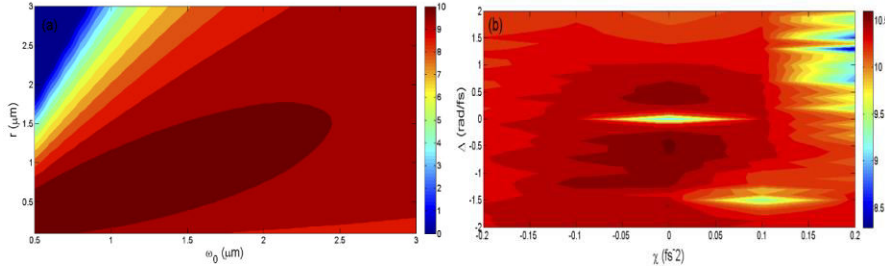

Fig. 3 Contour plot of $log_{10}(|acceleration/g|)$ as a function of (a) $r$ and $\omega_0$ (b) $\chi$ and $\Delta$ under the same pulse parameters as in Fig. 2 and $\rho_{22}(-\infty) = 1$.

Fig.3 depicts the role of the various laser pulse parameters on the acceleration produced by the optical force. Fig. 3(a) shows acceleration about 9 to 10 orders of magnitude higher than $g$ can be obtained as long as $r$ remains comparable or less than $\omega_0$, however it drops significantly as $r$ increases over $\omega_0$ and become almost zero for larger $r$ and smaller $\omega_0$. While Fig. 3(b) shows that the order of acceleration varies only between 10 to 10.5 over most of the region of $\chi$ and $\Delta$ for specific $r$ and $\omega_0$, exhibiting the robustness of the scheme over detuning frequency and chirp rate.

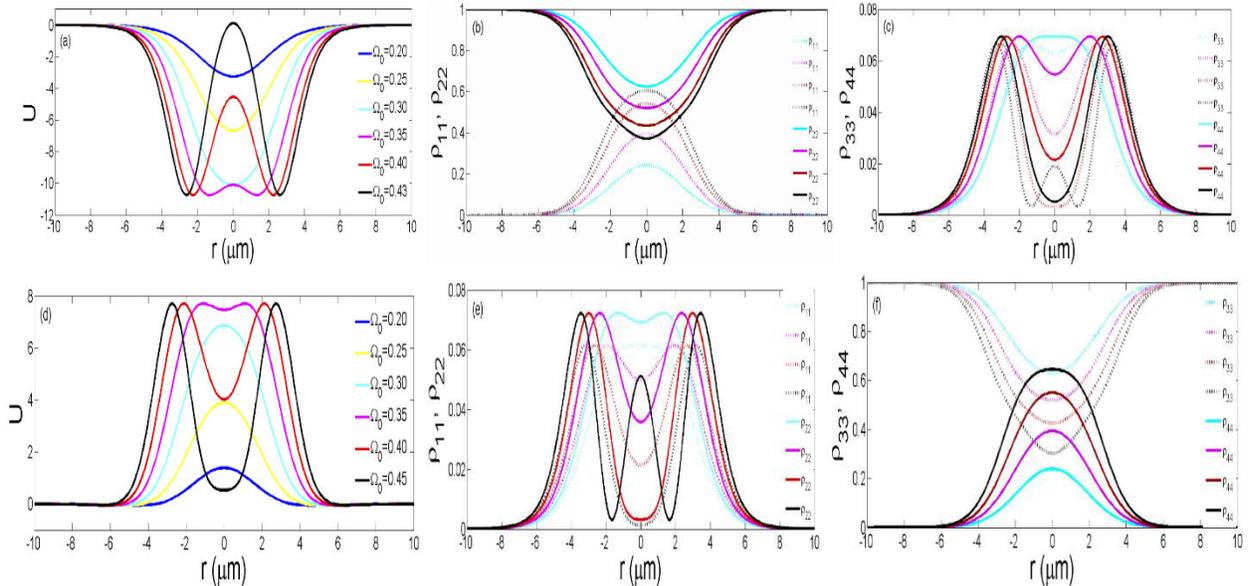

FIG. 4. (a) Spatial profile of optical dipole potential under $\rho_{22}(-\infty) = 1$, $\omega_0 = 5\mu m$, other parameters are same as Fig. 2 (b) Spatial probability distribution (SPD) of $\rho_{11}$ (dotted) and $\rho_{22}$ (solid) (c) $\rho_{33}$ (dotted) and $\rho_{44}$ (solid). The different colors correspond to the same Rabi frequency as in Fig. 4(a). (d) Spatial profile of optical dipole potential with $\rho_{33}(-\infty) = 1$ (e) SPD of $\rho_{11}$ (dotted) and $\rho_{22}$ (solid) (f) SPD of $\rho_{33}$ (dotted) and $\rho_{44}$ (solid). The different colors correspond to the same Rabi frequency as in Fig. 4(d).

We may be able to derive more information about the nature of optical force by studying optical potentials. So next, we study the effect of various controllable parameters such as Rabi frequency, beam waist and detuning frequency on optical dipole potential. Fig. 4(a) and (d) depicts the spatial profile of the optical dipole potential subject to the initial populations in level $|2>$ and $|3>$ respectively for various Rabi frequencies. It can be observed from Fig. 4(a) that there is no splitting in the optical potential up to $\Omega_0 = 0.30$ rad/fs. However, when $\Omega_0 = 0.43$ rad/fs the potential splits into two about the center of the trap potential. Similar behavior is observed in Fig. 4(d) also. This trap splitting phenomenon matches quite well with the recent experimental work by Y. Jiang *et al* on trapping of nanoparticles with femtosecond pulses [12]. In order to understand the physics of the problem, it may be useful to know the spatial probability distribution of different levels with increasing Rabi frequency. Fig. 4(b) and 4(e) shows the spatial probability distribution for the two ground levels, corresponding to Fig. 4(a) and 4(d) respectively. On the other hand, Fig. 4(c) and 4(f) depict the same for the two excited levels. Fig. 4(b) shows that with increasing Rabi frequency, the maximum and minimum value of $\rho_{11}$ and $\rho_{22}$ increases, respectively. Fig.4(c) reveals that the spatial probability distribution of the two excited levels get split with increasing $\Omega_0$. On the other hand corresponding to Fig. 4(d), Fig. 4(e) refers to the splitting of the ground levels and Fig. 4(f) represents the increase of maximum and minimum value for $\rho_{33}$ and $\rho_{44}$ respectively. So in the femtosecond time scale, with increasing Rabi frequency, the spatial probability distribution of different levels get split depending on the initial population of the atom, lead to the splitting of optical potential. Fig. 5 depicts the optical dipole potential for different beam waists. It can be seen that the spatial size of the trap decreases with decreasing beam waist of the pulses. It should be noted that we have kept $\Omega_0 = 0.30$ rad/fs to avoid splitting. So one can have fine control over the trap size by tuning the beam waists of the pulse.

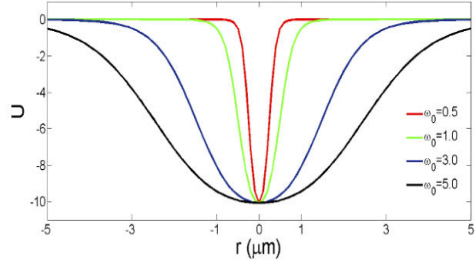

FIG. 5. Spatial profile of optical dipole trap with different beam waist and $\rho_{22}(-\infty) = 1$ $\Delta = -0.7\ rad/fs$, $\chi = 0.01\ fs^{-2}$ and $\Omega_0 = 0.30\ rad/fs$.

Finally in Fig. 6 we depict the spatiotemporal profile of the optical dipole potential for different detuning frequencies. The optical potential is negative around the z-axis. So a four level atom can be trapped by this time dependent potential. We observe that as the detuning frequency decreases, the number of minima in the potential decreases and finally when the detuning frequency is $\Delta = -0.2\ rad/fs$ there is only a single trap over the entire region. Thus, the trapping region for atoms can be well controlled by judiciously choosing the detuning frequency.

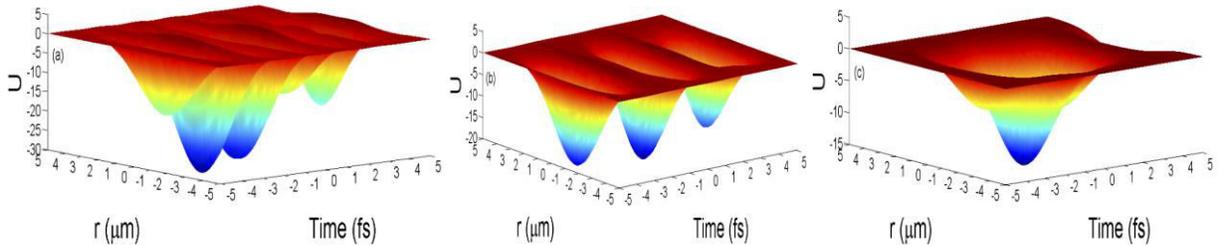

FIG. 6. Spatiotemporal profile of optical potential trap with $\rho_{22}(-\infty) = 1$, $\chi = 0.01\ fs^{-2}$, $\Omega_0 = 0.30\ rad/fs$ and $\omega_0 = 3\mu m$. (a) $\Delta = -0.7\ rad/fs$, (b) $\Delta = -1.5\ rad/fs$ and (c) $\Delta = -2.0\ rad/fs$

In conclusion we have studied optical dipole force and potential on a N-type four level atomic system. It is observed that the transverse optical dipole force changes from focusing to defocusing one as the initial population changes from ground levels to the excited levels. The atoms are accelerated by 10 orders of magnitude higher than the Earth's gravitational acceleration $g$. The dependency of acceleration of an atom on off-axis position, the beam waist, the detuning frequency and chirping rate is also studied. The spatial size of the trap can be well controlled by tuning the beam waist of the Gaussian pulse and the detuning frequency. The trapping potential splits with increasing Rabi frequency about the center of the trap, a behavior analogous to the one observed experimentally in the context of trapping of nanoparticles with femtosecond pulses. This work may trigger more systematic studies in the direction of trapping and manipulation of atoms and molecules.

**Acknowledgements**

Authors would like to acknowledge the financial support from CSIR, India [Grant No. 03(1252)/12/EMR-II].